\documentclass[runningheads]{llncs}
\usepackage{graphicx}
\usepackage{amsmath}

\begin{document}
\title{Multi-task Localization and Segmentation for X-ray Guided Planning in Knee Surgery}
\titlerunning{Multi-task Localization and Segmentation for Surgical Planning}

\author{Florian Kordon\inst{1,2,3} \and
Peter Fischer\inst{1,2} \and
Maxim Privalov\inst{4} \and
Benedict Swartman\inst{4} \and
Marc Schnetzke\inst{4} \and
Jochen Franke\inst{4} \and
Ruxandra Lasowski\inst{3} \and
Andreas Maier\inst{1} \and
Holger Kunze\inst{2}}

\authorrunning{Kordon et al.}
\institute{
Pattern Recognition Lab, Department of Computer Science, Friedrich-Alexander-Universität Erlangen-Nürnberg, Erlangen, Germany \\ \email{florian.kordon@fau.de} \\ \and
Advanced Therapies, Siemens Healthcare GmbH, Forchheim, Germany 
\\ \and
Faculty of Digital Media, Hochschule Furtwangen, Furtwangen, Germany \\ \and
Department for Trauma and Orthopaedic Surgery, BG Trauma Center Ludwigshafen, Ludwigshafen, Germany}
\maketitle              
\begin{abstract}
X-ray based measurement and guidance are commonly used tools in orthopaedic surgery to facilitate a minimally invasive workflow. 
Typically, a surgical planning is first performed using knowledge of bone morphology and anatomical landmarks. Information about bone location then serves as a prior for registration during overlay of the planning on intra-operative X-ray images. Performing these steps manually however is prone to intra-rater/inter-rater variability and increases task complexity for the surgeon.
To remedy these issues, we propose an automatic framework for planning and subsequent overlay. We evaluate it on the example of femoral drill site planning for medial patellofemoral ligament reconstruction surgery. A deep multi-task stacked hourglass network is trained on 149 conventional lateral X-ray images to jointly localize two femoral landmarks, to predict a region of interest for the posterior femoral cortex tangent line, and to perform semantic segmentation of the femur, patella, tibia, and fibula with adaptive task complexity weighting.
On 38 clinical test images the framework achieves a median localization error of 1.50 mm for the femoral drill site and mean IOU scores of 0.99, 0.97, 0.98, and 0.96 for the femur, patella, tibia, and fibula respectively. 
The demonstrated approach consistently performs surgical planning at expert-level precision without the need for manual correction. 

\keywords{Landmark localization \and Multi-label bone segmentation \and MPFL \and X-ray guidance \and Orthopaedics \and Surgical planning.}
\end{abstract}

\section{Introduction}
\begin{figure}[b]
\includegraphics[width=\textwidth]{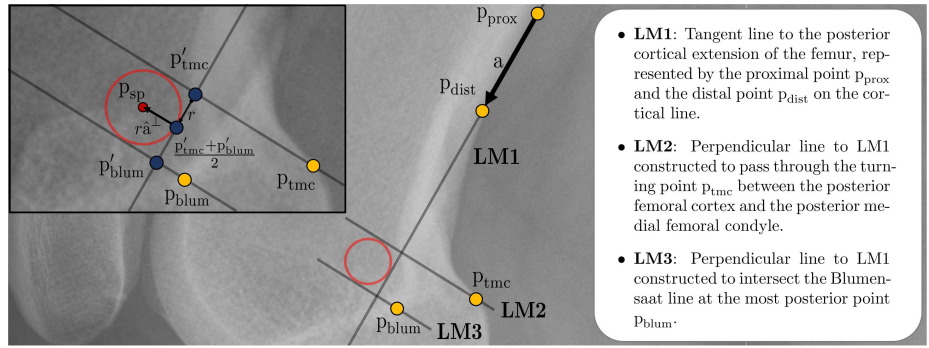}
\caption{Approximation of the Schoettle Point $\textrm{p}_{\textrm{sp}}$~\cite{Schoettle.2007} as the center of the inner circle of three lines. These lines can be derived from osseous landmarks on a lateral radiograph.} \label{fig:derivation}
\end{figure}

In orthopaedics, X-ray imaging is frequently used to facilitate planning and operative guidance for surgical interventions. By capturing patient-specific characteristics and contextual information prior to and during the procedure, such image-based tools benefit a more reliable and minimally invasive workflow at reduced risk for the patient. To this end, typical assessment involves geometric measurements of patient anatomy, verification of correct positioning of surgical tools and implants, as well as navigational guidance with help of anatomical landmarks and bone morphology.
In current clinical practice, several methodologies have been established which leverage this toolset to standardize routine procedures. One example is the Schoettle planning methodology for reconstruction surgery of a ruptured medial patellofemoral ligament (MPFL)~\cite{Schoettle.2007}. To restore the anatomically correct biomechanics and to forestall recurrent injuries, the optimal fixation area on the femur is approximated by the Schoettle Point, which can be derived from several osseous landmarks (Fig.~\ref{fig:derivation}). 
Unfortunately, execution of such a methodology faces several clinical and technical challenges~\cite{Joskowicz.2016,Szekely.2016}. First, many orthopaedic surgeries target anatomical regions which are not directly inferable from the image but rely on auxiliary structures derived from anatomical landmarks, leading to inter-rater and intra-rater differences. 
Secondly, the overlay of the planning result on subsequent intra-operational images requires registration to compensate for motion which should be restricted to the anatomical region of interest (ROI), in the case of MPFL, the femoral bone. 
And lastly, manual intra-operational planning and interaction with a guidance application in a sterile setting are disruptive in the doctor's surgical workflow.

Using MPFL reconstruction as an example, we present a framework which allows fully-automatic localization of anatomical landmarks, semantic segmentation of bone structures, and prediction of ROIs for geometric line features on X-ray images. Building upon the ideas of~\cite{Bier.2018}, we exploit recent advances in sequential deep learning architectures in form of deep stacked hourglass networks (SHGN)~\cite{Newell.2016} to refine predictions based on the learned residual information between the ground truth and intermediate estimates. We propose an extension to a multi-task learning approach to incorporate cross-task information for an enriched and more general feature set, which proved to be beneficial in X-ray based segmentation tasks~\cite{Breininger.2018}. To automatically weight the single task loss terms, our framework introduces a novel adaption of gradient normalization~\cite{Chen.2018} for stacked network architectures by integrating it with a deeply supervised optimization scheme.
We evaluate this approach for femoral attachment site planning in MPFL reconstruction surgery which is a clinically relevant and common procedure. We demonstrate expert-level performance of our proposed solution with a comprehensive evaluation including clinical data and an inter-rater study with multiple surgeons.
The achieved results enable direct integration into the operative workflow and in almost all cases allow the number of manual planning steps to be limited to the confirmation of the planning proposal, so that the surgeon can remain sterile throughout the procedure. 

\section{Methods}
\subsection{Multi-task Stacked Hourglass Network}
\begin{figure}[hb]
\includegraphics[width=\textwidth]{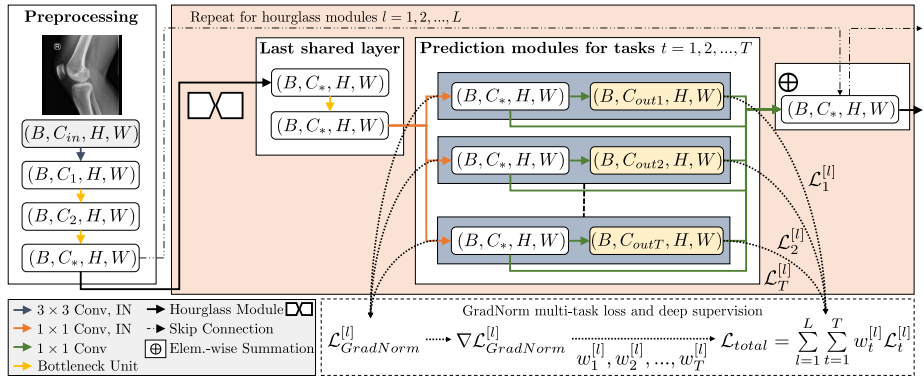}
\caption{Proposed multi-task network based on the SHGN architecture with intermediate GradNorm weight balancing, instance normalization (IN), and pre-activation layout for residual bottleneck units~\cite{Chen.2018,He.2016,Newell.2016}. Here, $C_{in}=1$, $C_1=64$, $C_2=128$, and $C_*=256$.} \label{fig:architecture}
\end{figure}

A SHGN is a multi-stage convolutional network architecture which sequentially arranges $l=1,2,...,L$ symmetrical Fully Convolutional Networks referred to as hourglass modules (HG)~\cite{Newell.2016}. By cascaded inference,
several iterations of bottom-up and top-down processing of data and features are performed to capture and combine the input morphology at various scales and abstraction levels. At the end of the expanding path of each HG, features are fed into an additional bottleneck residual unit before being distributed for individual task processing. For each task $t=1,2,...,T$ we introduce a separate prediction module to facilitate task-specific discriminative power, allow for intermediate estimates, and exploit iterative refinement by reinjection (Fig.~\ref{fig:architecture}).

\subsubsection{Weighted Multi-label Segmentation Loss}
X-ray images are superimposition projections, which leads to ambiguities in class assignment for overlapping bones with similar imaging characteristics. 
Instead of using multinomial pixel-wise classification, we therefor define bone segmentation as a multi-class/multi-label problem and perform separate binomial classifications for each bone in the target region to allow a pixel to be assigned multiple labels. We further exploit this multi-label information to penalize errors in overlap regions, which we derive by a characteristic function $g_{bij} = \left[\sum_{c}y_{bcij} > 1 \right]$ and incorporate into the loss function with scaling factor $\beta$. $y$ corresponds to a 4th-order tensor $(B,C,H,W)$, where the task-specific ground truth maps are stacked along $C$. Each tensor element is indexed with $b \in [1,B]$, $c \in [1,C]$, $i \in [1,H]$, and $j \in [1,W]$ where $B$, $C$, $H$, $W$ mark batch, channel, height, and width dimensions. The resulting segmentation loss for prediction $\hat{y}^{[l]}$ with sigmoid nonlinearity $\sigma$ computes by  
{\thinmuskip=0mu\medmuskip=0mu\thickmuskip=0mu
\begin{equation}
\mathcal{L}_{\text{seg}}^{[l]} = \frac{1}{BHW}
\sum_{(b,c,i,j)}
\left( 1+\beta g_{bij} \right)
\left[  
- y \log \left( \sigma 
\left( \hat{y}^{[l]} \right) \right) 
- (1-y) \log \left( 1- \sigma \left( \hat{y}^{[l]} \right) \right)
\right]_{bcij}.
\end{equation}}

\subsubsection{Landmark and Region of Interest Loss}
Landmarks and ROIs are represented as heat-maps, which encode the localization likelihood as a spatial intensity distribution. For landmark ground truth, a 2D unnormalized Gaussian with a standard deviation of $6$ pixels is centered on the annotated position. Likewise, the line's ROI ground truth is derived by placing equidistant pseudo landmarks along the ground truth cortical line.
For loss calculation, heat-map matching is performed as 
\begin{equation}
\mathcal{L}_{\{lm, roi\}}^{[l]} = \frac{1}{BCHW} \sum_{(b,c,i,j)} \left({\hat{y}_{bcij}^{[l]}-y_{bcij}} \right)^2.
\end{equation}
We derive landmark positions by performing the $\textrm{arg}\,\textrm{max}$ operation on the predicted likelihood scores. For the posterior femoral cortex tangent line LM1 (Fig.~\ref{fig:derivation}), we mask the femur segmentation outline with the predicted ROI and discard features with a likelihood below $0.5$ prior to a least squares regression. 

\subsubsection{Task Weighting and Total Loss}
We utilize deep supervision to ease gradient flow across multiple stages of the SHGN and to facilitate faster training. For this purpose, individual task losses are calculated and summed up for both intermediate and final HGs to form a single loss value.
To respect imbalanced task difficulties and to avoid overfitting to only a subset of tasks, we use gradient normalization (GradNorm) and adapt it to a deeply supervised setting~\cite{Chen.2018}. GradNorm generally tackles task imbalance by reducing the variance across the tasks' training rates. For this purpose, task-specific loss weighting factors are learned by jointly reducing an additional multi-task loss function on the basis of gradient magnitudes to adaptively adjust the gradient norm at each update step~\cite{Chen.2018}. The GradNorm weights for each supervised HG are based on the last shared bottleneck layer before branching off to the prediction modules (Fig.~\ref{fig:architecture}).
The resulting balanced loss is given by $\mathcal{L}_{total} = \sum_{l=1}^{L} w_{seg}^{[l]} \mathcal{L}_{seg}^{[l]} + w_{lm}^{[l]} \mathcal{L}_{lm}^{[l]} + w_{roi}^{[l]} \mathcal{L}_{roi}^{[l]}$.

\subsection{Dataset and Training Procedures}
Training and validation data consists of 185 lateral X-ray projections of the knee joint acquired prior to reconstruction surgery. The data was split with ratio $0.8/0.2$ for training and validation (149/36 images).  
For evaluation, 38 separate test images with standardized measuring spheres of 30 mm diameter were used. Annotation of the ground truth landmark positions and line reference points on training and validation images was performed by one orthopaedic surgeon with an interactive proprietary tool (Fig.~\ref{fig:derivation}). 
To allow for an estimate on inter-rater variability, annotation on the test data was extended to three orthopaedic surgeons from the same hospital. Ground truth segmentation masks for the femur, patella, tibia, and fibula were created by the first author.
A basic set of data augmentations (rotation, scaling, horizontal flipping) as well as linear contrast scaling with a probability of $p = 0.5$ each were applied during training. After augmentation, the variably sized images were zero-padded to square spatial dimensions and subsequently downsampled to a resolution of $256 \times 256$ pixels. 

We devise a multi-task SHGN with $L=4$ HGs and introduce instance normalization layers for approximate contrast invariance and to smooth the optimization landscape~\cite{Kordon.2019}. We consider $T=3$ tasks and hence construct three prediction modules at the end of each HG. The network was implemented with PyTorch (v0.4.1) and trained on a NVIDIA Quadro P5000 over 250 epochs with batch size 2. The network parameters and GradNorm task weights were optimized with RMSProp at learning rates of 0.00025 and 0.025 respectively. The learning rate for network parameters was halved every 60 epochs. Based on prior hyper-parameter optimization, GradNorm's asymmetry hyper-parameter was set to $\alpha=1$ at each HG, and the penalizing weight factor for multi-label segmentation was assigned to $\beta=0.6$.

\section{Evaluation}

\subsubsection{Bone Segmentation}
The model consistently yields high overlap- and contour-based metric results and successfully delineates all target bone structures (Tab.~\ref{tab:segmentation}). Qualitative assessment indicates successful disambiguation in overlapping areas, in narrow interarticular joint spaces, and in low-contrast regions (Fig.~\ref{fig:combresults}). Also, uncommon image characteristics like osteophytes along joint contours as well as aberrant lateral projections are resolved with high precision. However, subpar performance is observed for the fibula due to the proximal part being mostly overlapped by the tibia with seemingly no intensity shifts. Likewise, wrongful assignment of spherical markers to the tibia or the femur leads to high contour distances. 

\begin{table}[htb]
\centering
\caption{Segmentation performance on 38 test images for all bones in target region.}\label{tab:segmentation}
\begin{tabular}{|l|c|c|c|}
\hline
Anatomy & \begin{tabular}[c]{@{}c@{}}mean IOU \\ (Mean $\pm$ STD)\end{tabular} & \begin{tabular}[c]{@{}c@{}}Average Surface Distance \\ (Mean $\pm$ STD) (mm)\end{tabular} & \begin{tabular}[c]{@{}c@{}}Hausdorff Distance\\ (Mean $\pm$ STD) (mm)\end{tabular} \\ \hline
Femur & $0.99\pm0.01$ & $0.12\pm0.61$ & $2.96\pm7.51$ \\
Patella & $0.97\pm0.02$ & $0.02\pm0.02$ & $0.62\pm0.56$ \\
Tibia & $0.98\pm0.02$ & $0.23\pm0.85$ & $3.76\pm10.77$ \\
Fibula & $0.96\pm0.02$ & $0.14\pm0.68$ & $2.38\pm5.41$ \\ \hline
\end{tabular}
\end{table}

\subsubsection{Line and Landmark Localization}
\begin{figure}[tb]
\centering
\includegraphics[width=\textwidth]{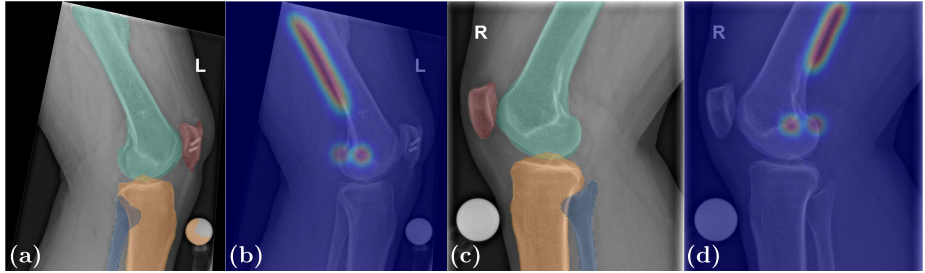}
\caption{Automatic results for multi-task segmentation (a,c) and localization (b,d). In (a), false-positive assignment of a spherical marker to the tibia is observed.} \label{fig:combresults}
\end{figure}

Predictions for the landmarks $p_{\textrm{blum}}$ and $p_{\textrm{tmc}}$ are spatially precise with median Euclidean distance (ED) errors of $1.18,\allowbreak\,\textrm{CI}_{80\,\%}[0.99,1.74]\,\textrm{mm}$ and $2.14,\,\textrm{CI}_{80\,\%}[1.71,2.63]\,\textrm{mm}$ respectively (Fig.~\ref{fig:combresults}). In general, it can be observed that localization of $p_{\textrm{tmc}}$ is less robust due to its dependence on true-lateral imaging. Slight deviations from a true-lateral projection lead to non-overlapping femoral epicondyles, which necessitates three-dimensional reasoning and compensation for correct spatial positioning. For measuring the alignment of the cortical extension line, ED of the ground truth points $p_{\textrm{prox}}$ and $p_{\textrm{dist}}$ to the predicted line are averaged, yielding a median score of $0.62,\,\textrm{CI}_{80\,\%}[0.48,0.79]\,\textrm{mm}$.

\subsubsection{Adaptive Task Weighting}
\begin{figure}[tb]
\includegraphics[width=\textwidth]{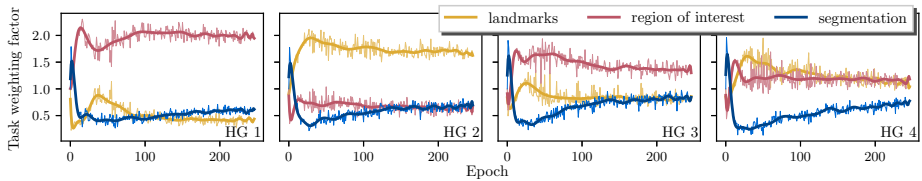}
\caption{Development of GradNorm task weights for each HG during training.} \label{fig:task-weights}
\end{figure}

The learned GradNorm task weights generally reduce the segmentation training rate across all modules in exchange for increased landmark and ROI loss contributions (Fig.~\ref{fig:task-weights}). With advanced training time, balancing slightly converges which indicates harmonization of the task-specific loss magnitudes and gradients. Especially in early HGs, optimization towards a single task is observed. 

\subsubsection{Inter-rater Analysis and Automatic Planning}
\begin{figure}[t]
\includegraphics[width=\textwidth]{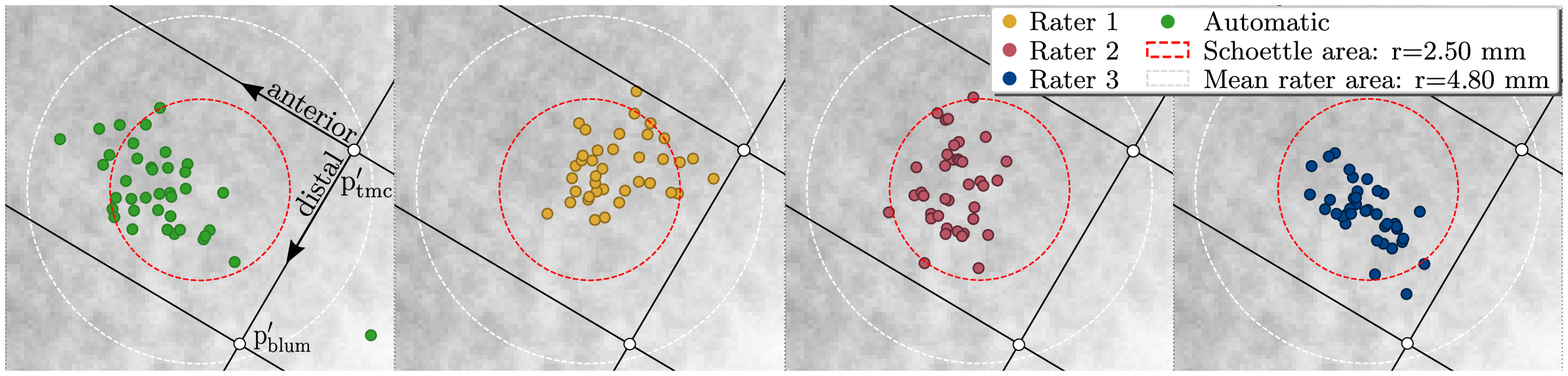}
\caption{
Visualization of errors between individually planned Schoettle Points and the experts' centroid. Distance scores are standardized w.r.t. pixel/mm spacing and to the mean orientation of LM1 based on expert annotations. 
Per image, all results are visually aligned to a reference planning, whose Schoettle Point corresponds to this centroid. For comparison with the original Schoettle area of $r=2.50~\textrm{mm}$, the average confidence circle as planned by the experts (bounded by LM1, LM2 and LM3) is overlayed.
}\label{fig:interrater}
\end{figure}

The observed inter-rater EDs of the constructed Schoettle Points are generally within a circular confidence area with radius $r = 2.50\,\textrm{mm}$, which describes an anatomically correct femoral MPFL insertion site~\cite{Schoettle.2007}. In comparison, automatic plannings are equally reliable and can reduce variability caused by differences in planning strategy between expert raters (Tab.~\ref{tab:interrater}, Fig.~\ref{fig:interrater}). However, morphological variations of the femoral cortex or a too short predicted line ROI frequently lead to slight anterior shifting. Likewise, projection of individual ED errors onto the longitudinal and anteroposterior axes indicates an error tendency towards the anterior direction, which underlines difficulties in correct assessment of the femoral bow (Fig.~\ref{fig:interrater}).

\begin{table}[htb]
\centering
\caption{Inter-rater variability and comparison with proposed automatic planning. The median ED at 80 \% confidence level (mm) between raters is reported for full dataset and subset of images agreed to be suitable for surgical planning by all raters.
}\label{tab:interrater}
\begin{tabular}{|l|l|c|c|}
\hline
First rater & Second rater & \begin{tabular}[c]{@{}c@{}}Schoettle Point\\ (38/38 test images)\end{tabular} & \multicolumn{1}{l|}{\begin{tabular}[c]{@{}c@{}}Schoettle Point\\ (29/38 suitable test images)\end{tabular}} \\ \hline
1 & 2 & 2.35, {[}1.94, 2.85{]} & 2.68, {[}2.09, 3.13{]} \\
1 & 3 & 2.31, {[}1.91, 2.79{]} & 2.49, {[}1.91, 2.95{]} \\
2 & 3 & 1.67, {[}1.37, 2.22{]} & 1.62, {[}1.07, 2.12{]} \\
Autom. & 1 & 2.41, {[}1.97, 2.99{]} & 2.64, {[}1.64, 3.10{]} \\
Autom. & 2 (gr. truth training) & 1.46, {[}1.00, 1.85{]} & 1.33, {[}0.91, 1.59{]} \\
Autom. & 3 & 1.61, {[}1.45, 1.87{]} & 1.56, {[}1.27, 1.73{]} \\
Autom. & Expert centroid & 1.50, {[}1.41, 2.07{]} & 1.41, {[}1.28, 1.52{]} \\ \hline
\end{tabular}
\end{table}

\section{Discussion and Conclusion}
We proposed an automatic framework for joint prediction of segmentations and heatmaps for spatial localization of landmarks and line features. On the example of MPFL reconstruction surgery, we could show that we can facilitate surgical planning by providing planning proposals at expert-rater precision. We see limitations in that the proposed method was only trained and tested on pre-operational, conventional X-ray images which depict a large portion of the femoral shaft and typically have high contrast. For seamless integration into a clinical workflow with subsequent overlay of the planning result on live images, the framework performance must be evaluated on fluoroscopic image data. This modality however imposes additional difficulties onto automated prediction by superimposed surgical tools and by greater overall heterogeneity in image characteristics and acquisition settings. This could partially by solved by overlay of simulated tools and implants in the image domain during training, which showed to increase robustness of learning-based algorithms~\cite{Kordon.2019}.  

As shown in this work, planning methods like the Schoettle methodology might also inherently tolerate variability in assessment strategy and typically cannot be securely validated due to absence of anatomical ground truth. An automatic solution should therefor utilize an adequate encoding of this variability to alleviate overfitting to a certain type of annotation strategy by a single rater. Furthermore, while we experience satisfactory results in estimating a line by masking segmentation contour features, such cross-task coupling introduces additional failure points in the planning pipeline. As for estimation of the posterior femoral cortex, strongly curved femoral shafts allow an anterior shift of the fitted line, which is directly conditioned by the longitudinal extends of the predicted ROI. To this end, we aim to look at ways to derive confidence estimates for each task and to keep the number of interdependent tasks at a reasonable level. In future work, we also seek to adapt our approach to different anatomies by exploiting the highly generalizable concept of heat-map matching for a direct representation of arbitrarily shaped features. 

\subsubsection{Disclaimer} The methods and information presented here are based on research and are not commercially available.

\bibliographystyle{splncs04}
\bibliography{ms}
\end{document}